\documentstyle[aps,twocolumn]{revtex}
\input{psfig}
\begin{document}
\draft
\title{Frequency Dependence of Magnetopolarizability of Mesoscopic
Grains}  
\author{Ya.~M.~Blanter$^{a}$ and A.~D.~Mirlin$^{b,c,*}$}
\address{$^a$ D\'epartement de Physique Th\'eorique, Universit\'e de
Gen\`eve, CH-1211 Gen\`eve 4, Switzerland\\
$^b$ Institut f\"ur Nanotechnologie, Forschungszentrum Karlsruhe,
76021 Karlsruhe, Germany\\
$^c$ Institut f\"ur Theorie der Kondensierten Materie,
Universit\"at Karlsruhe, 76128 Karlsruhe, Germany}
\date{\today}
\maketitle 
\tighten
\begin{abstract}
We calculate average magnetopolarizability of an isolated metallic
sample at frequency $\omega$ comparable to the mean level spacing
$\Delta$. The frequency dependence of the magnetopolarizability is
described by a universal function of $\omega/\Delta$. 
\end{abstract}

Mesoscopic effects in electric polarizability of small metallic
particles have been discussed in the literature starting from the
seminal work by Gor'kov and Eliashberg (GE) \cite{GE}. The role of
screening effects, which were not taken into account in the original
paper \cite{GE}, has been subsequently emphasized \cite{ABB1}. Whereas
GE predicted a giant effect of level correlations on the
polarizability, screening actually reduces this quantum effect to a
relatively small correction to the classical value. Like other
mesoscopic effects related to quantum interference, this correction is
affected by magnetic field. It can thus be observed in
magnetopolarizability of a grain, similarly to the weak localization
correction to resistivity \cite{Ef1,Noat,BM}. An experimental study of 
quantum corrections in electromagnetic response of an ensemble of
$10^5$ mesoscopic samples has been reported recently
\cite{Noat1,deblock00}.  

The aim of this paper is to calculate the dependence of the
magnetopolarizability on the frequency $\omega$ of the applied
electric field in the range $\omega\sim\Delta$, where $\Delta$ is the
mean level spacing. This dependence is given by a {\em universal}
function of the parameter $\omega/\Delta$. Note that the frequency in
the experiments \cite{Noat1,deblock00} was comparable to $\Delta$, and
thus experimental verification of this scaling form is in principle
feasible.   
  
Previously we have demonstrated \cite{BM1,BM} how the supersymmetric
sigma-model can be employed to calculate magnetopolarizability in
the grand canonical ensemble. However, the experimentally relevant
situation of isolated grains is appropriately described in the
framework of the canonical ensemble. The crucial role of the
statistical ensemble for the problem considered has been emphasized in
Ref. \cite{shkl} (see also Ref. \cite{Gefen}). Technically,
calculation of the dynamical response within the canonical ensemble is
more complicated, since it involves three-level correlation functions, 
which poses a serious obstacle to the application of the supersymmetry
technique. Below, we bypass this difficulty by combining information
on statistical properties of wavefunctions, obtained with the
supersymmetry technique, with the three-level correlation function
known from the random matrix theory (RMT). 

The (complex-valued) polarizability is defined as $\bbox{d} (\omega) =
\alpha (\omega) \bbox{E} (\omega)$, where $\bbox{E}$ and $\bbox{d}$
are the external electric field and the induced dipole moment,
respectively. The expression for the ensemble-averaged
magnetopolarizability $\alpha_B (\omega) \equiv \alpha (\omega, B) -
\alpha (\omega, 0)$, where $B$ is an applied magnetic field (strong
enough to break the time-reversal symmetry) has the form \cite{Ef1,BM}  
\begin{eqnarray} \label{gen1}
\alpha_B (\omega) = \frac{2e^2}{E^2} \int d\bbox{r_1} d\bbox{r_2}
\Phi (\bbox{r_1}) \bbox{\delta} \Pi(\bbox{r_1}, \bbox{r_2};\omega)
\Phi (\bbox{r_2}),
\end{eqnarray}
where $\Phi(\bbox{r})$ is the local potential in the sample
(resulting from screening of the external electric field), the
symbol $\bbox{\delta}$ denotes the difference between quantities with
and without magnetic field, and $\Pi$ is the polarization operator,  
\begin{eqnarray} \label{gen2}
\Pi(\bbox{r_1}, \bbox{r_2};\omega) & = & \left\langle\sum_{m \ne n}
\psi_m^*(\bbox{r_1}) 
\psi_n (\bbox{r_1}) \psi_n^*(\bbox{r_2}) \psi_m (\bbox{r_2}) 
\right.\nonumber
\\
& \times & \frac{n_F(\epsilon_m) - n_F(\epsilon_n)}{\omega -
\epsilon_m + \epsilon_n}\Biggr\rangle.  
\end{eqnarray}
Here $m$ and $n$ label  exact single-particle states, the angular
brackets denote the ensemble averaging, and it is assumed that
$\omega$ has an (infinitesimally small) positive imaginary part,
$\omega\equiv\omega+i0$. We will consider the low-temperature limit,
$T\ll\Delta$, thus setting $T=0$ in the sequel. Substitution of
(\ref{gen2}) into (\ref{gen1}) yields 
\begin{eqnarray} \label{gen3}
\alpha_B (\omega) &=& \frac{2e^2}{E^2} \bbox{\delta} \left\langle
\sum_{\epsilon_n < \epsilon_F < \epsilon_m}  \left\vert 
\Phi_{mn} \right\vert^2 \right. \nonumber \\
&\times& 
\left(\frac{1}{\epsilon_m - \epsilon_n -\omega}+
\frac{1}{\epsilon_m - \epsilon_n +\omega}\right)
\Biggr\rangle,   
\end{eqnarray}
where $\Phi_{mn} = \int d\bbox{r} \psi_m^* (\bbox{r}) \Phi(\bbox{r})
\psi_n (\bbox{r})$. 

The canonical ensemble is realized by pinning the Fermi-level to one
of the single-particle levels $\epsilon_k$: $\epsilon_F = \epsilon_k +
0$. Splitting the sum in (\ref{gen3}) into  two contributions with
$n=k$ and $n\ne k$, we find
\begin{eqnarray} \label{ce1}
\alpha_B & = & \frac{2e^2}{E^2\Delta^2} \bbox{\delta} \left\{
\int_{+0}^{\infty} d\epsilon \left({1\over
\epsilon-\omega}+{1\over\epsilon+\omega}\right)  \right. \\
& \times & \left[ R_2(\epsilon) \Delta
 +   \left. \int_{+0}^{\epsilon-0} d\epsilon_1 R_3(\epsilon,
\epsilon_1) \right] \left\langle |\Phi|^2 
\right\rangle_{\epsilon} \right\}\ , \nonumber
\end{eqnarray}
where $R_2(\epsilon)$ and $R_3(\epsilon,\epsilon_1)$ are the two-level
and the three-level correlation functions (normalized to unity at
$\epsilon,\epsilon_1\gg\Delta$), 
\begin{eqnarray}
\label{r2r3}
R_2(\epsilon) &=& \Delta^2\left\langle\sum_{ij}
\delta(E-E_i)\delta(E+\epsilon-E_j)\right\rangle\ ; \nonumber \\
R_3(\epsilon,\epsilon_1) &=& \Delta^3\left\langle\sum_{ijk}
\delta(E-E_i)\delta(E+\epsilon-E_j) \right.\nonumber \\
& \times & \delta(E+\epsilon_1-E_k)\Biggr\rangle\ ,
\end{eqnarray}
and $\left\langle |\Phi|^2  \right\rangle_{\epsilon}$ is the
average squared matrix element, 
\begin{eqnarray}
\label{matrel}
\left\langle |\Phi|^2 \right\rangle_{\epsilon} &=&
R_2^{-1}(\epsilon)\Delta^2\left\langle\sum_{ij} |\Phi_{ij}|^2
\right. \nonumber \\
& \times & \delta(E-E_i)\delta(E+\epsilon-E_j)\Biggr\rangle\ .
\end{eqnarray}
When writing Eq.~(\ref{ce1}), we decoupled the wave function
correlations (\ref{matrel}) from the three-level correlation function
$R_3(\epsilon,\epsilon_1)$. Indeed, we know from the supersymmetry
calculations of two-level correlation functions \cite{EM} that (i) the
level correlation function has the RMT form, up to $1/g^2$
corrections; (ii) the wavefunction correlations have no dependence on
$\omega/\Delta$ in the order $1/g$. Here $g\sim
E_c/\Delta\gg 1$ is the dimensionless conductance of the grain, $E_c$
is the Thouless energy, and we consider the frequency range $\omega\ll
E_c$. We make thus an (extremely plausible) assumption that these
properties hold also for higher order correlation functions, which
allows us to proceed in the case of the canonical ensemble. 

We begin the evaluation of Eq.~(\ref{ce1}) by considering the term
which contains the three-level correlator. We write
$R_3(\epsilon,\epsilon_1)=R_2(\epsilon)
+\tilde{R}_3(\epsilon,\epsilon_1)$ and denote the corresponding
contributions as $\alpha_B^{(1)}$ and  $\alpha_B^{(2)}$. In the
leading order in $1/g$ the $\epsilon$-integral in $\alpha_B^{(1)}$ is 
a sum of contributions from the regions $\epsilon\sim E_c$ and
$\epsilon\sim\Delta$ (to be denoted as $\alpha_B^{(1a)}$ and
$\alpha_B^{(1b)}$ respectively). In the former we can neglect
$\omega$, which yields after taking into account the orthogonality and
completeness of the eigenfunctions \cite{Noat,BM}, 
\begin{eqnarray}
\label{alpha1a}
\alpha_B^{(1a)} &=& \frac{4e^2}{E^2\Delta^2} \bbox{\delta}
\int_{+0}^\infty d\epsilon R_2(\epsilon)
\left\langle |\Phi|^2 \right\rangle_{\epsilon} \nonumber \\
& = & -\frac{2e^2}{E^2\Delta} \left\langle |\Phi_{mm}|^2
\right\rangle = 
\frac{2e^2}{E^2\Delta} \left\langle |\Phi|^2
\right\rangle_0\ ,
\end{eqnarray}
where we defined
\begin{eqnarray}
\label{matrel0}
\left\langle |\Phi|^2 \right\rangle_0 & = & \left\langle |\Phi|^2
\right\rangle_{\epsilon\ll E_c} \\
& = & {1\over V^2} \int d\bbox{r_1} d\bbox{r_2}
\Phi(\bbox{r_1})\Phi(\bbox{r_2}) 
\Pi_D(\bbox{r_1}, \bbox{r_2})\ , \nonumber 
\end{eqnarray}
and $\Pi_D$ is the diffusion propagator satisfying
\begin{equation} \label{diff}
-D \nabla^2 \Pi_D (\bbox{r_1}, \bbox{r_2}) = (\pi\nu)^{-1} \left[
\delta(\bbox{r_1 - r_2}) - V^{-1} \right],
\end{equation}
with the  boundary condition $\nabla_{\bbox{n}} \Pi_D = 0$.
In the contribution $\alpha_B^{(1b)}$ we can neglect the
$\epsilon$-dependence of the matrix element, thus replacing it by 
$\left\langle |\Phi|^2 \right\rangle_0$,
\begin{eqnarray}
\label{alpha1}
\alpha_B^{(1b)} & = & \frac{2e^2}{E^2\Delta^2}
\left\langle |\Phi|^2 \right\rangle_0 \\
& \times & \int_{+0}^\infty d\epsilon
\left({1\over \epsilon-\omega} + {1\over \epsilon+\omega} \right)
\epsilon \bbox{\delta}R_2(\epsilon)\ . \nonumber
\end{eqnarray}
The same is valid for the term $\alpha_B^{(2)}$, as 
well as for the remaining contribution of the first term in square 
brackets in (\ref{ce1}).  Collecting everything, we finally get 
\begin{equation} 
\label{final} 
\alpha_B  =  \frac{2e^2}{E^2\Delta}  
\left\langle |\Phi|^2 \right\rangle_0 F(\omega)\ , 
\end{equation} 
where 
\begin{eqnarray} 
\label{f} 
F(\omega) &=& 1 + \int_{+0}^\infty {d\epsilon\over\Delta} 
\left({1\over \epsilon-\omega}+{1\over\epsilon+\omega}\right) \\ 
&\times & \left[\epsilon \bbox{\delta}R_2(\epsilon) + \Delta  
\bbox{\delta}R_2(\epsilon)  + \int_{+0}^{\epsilon-0} d\epsilon_1 
\bbox{\delta}\tilde{R}_3(\epsilon, \epsilon_1)\right]\ . \nonumber 
\end{eqnarray} 
Since the integrals in (\ref{f}) are determined by the range 
$\epsilon,\epsilon_1 \sim \Delta$, we can use the RMT results 
\cite{RMT} for the 
level correlation functions $R_2$ and $\tilde{R}_3$ entering this 
formula. Therefore, $F(\omega)$ is in fact a universal function of the 
dimensionless parameter $s=\omega/\Delta$. Real and imaginary parts of
this function describe the influence of the magnetic field on
polarizability and absorption of grains, respectively, and are plotted
in Fig.~1.   
\begin{figure}
\centerline{\psfig{figure=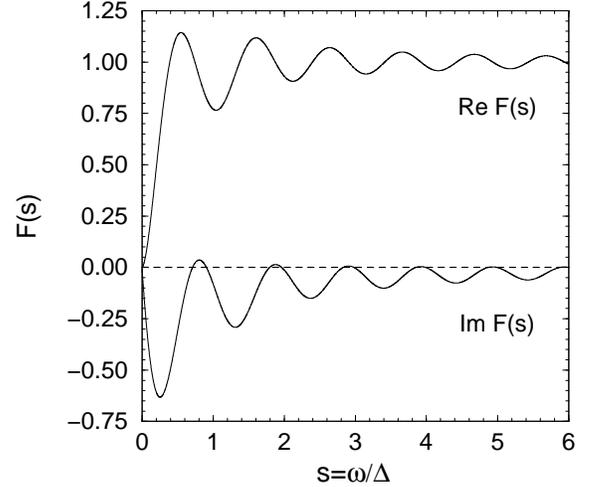,width=7.5cm}}
\vspace{0.3cm}
\caption{Real and imaginary parts of the function $F$
[Eq. (\protect\ref{f})] representing the frequency dependence of the
magnetopolarizability $\alpha_B(\omega)$.}  
\label{fig1}
 
\end{figure}
Note that $F(s)=0$ at $s=0$, which is an identity relating the
two-level and the three-level correlation functions entering
Eq.~(\ref{f}). This identity can be derived by using the
invariance of the level correlations with respect to a perturbation
\cite{footnote}. 

As is seen from Fig.~1, our results predict a positive
magnetopolarizability for all values of $\omega$ and a negative
magnetoabsorption for almost all $\omega$. This is in qualitative
agreement with the findings of Ref. \cite{deblock00}, where the
measurements have been performed at a frequency corresponding to
$s\simeq 0.21$. It would be very interesting to have the experimental
data for several frequencies in the range $\omega\sim \Delta$ in order
to check our prediction for the universal scaling function $F(s)$
describing the frequency dependence of the magnetopolarizability
induced by the level and eigenfunction statistics in the grains. 

We are grateful to H.~Bouchiat for stimulating discussions.
This work was supported by the Swiss National Science Foundation
(Y.~M.~B.) and the SFB 195 der Deutschen Forschungsgemeinschaft
(A.~D.~M.). We gratefully acknowledge the hospitality
of the Lorentz Center in Leiden, where this work was completed.

\end{document}